# MICROWAVE SURFACE RESISTANCE IN MGB$_2$.

A.A. Zhukov, K.Yates, G. K Perkins, Y. Bugoslavsky, M. Polichetti[+], A.Berenov, J.Driscoll, A.D. Caplin and L.F. Cohen

*Centre for High Temperature Superconductivity, Imperial College of Science Technology and Medicine, Prince Consort Road, London SW7 2BP, U.K.*

**Ling Hao and J. Gallop**

*National Physical Laboratory, Queens Rd, Teddington, UK*

**Abstract** Measurements of the temperature dependence of the surface resistance at 3 GHz of 100 micron size grains of MgB$_2$ separated powder are presented and discussed. The microwave surface resistance data is compared to experimental results of Nb, Bi$_2$Sr$_2$CaCu$_2$O$_{8+\delta}$ (BSCCO) and theoretical predictions of s-wave weak coupling electron-phonon theory (BCS).

**INTRODUCTION**

The recent discovery of superconductivity at 39 K in MgB$_2$ [1] has generated a great deal of interest. This is the highest reported superconducting transition temperature ($T_c$) in the class of bimetallic superconductors. The new superconductor appears to be hole-doped, as is evident from Hall effect measurements [2] and theoretical calculations [3-6]. Some initial results suggest that MgB$_2$ is a conventional superconductor with a single s-wave order parameter and intermediate or even strong phonon-mediated coupling $\lambda \sim 0.65-0.9$. This point of view is based on NMR measurements[7], the $^{11}$B isotope effect [8], neutron scattering [9], measurements of the shift of $T_c$ with hydrostatic pressure[10], specific heat measurements [11,12] and scanning tunnelling spectroscopy (STS) [13-17]. However, it is early days. The STS measurements in particular appear to have problems of reproducibility, most likely due to variations in surface quality, although there is some debate concerning gap anisotropy, multiple gaps etc [18].

The BCS model predicts activated behaviour i.e. $\exp(-\Delta/T)$. This behaviour should be reflected in the temperature dependence of the field penetration depth $\lambda(T)$. However, both linear [19], and quadratic [20] temperature dependence of $\lambda(T)$ have been reported. The former was extracted from $H_{c1}(T)$ (measured on bulk polycrystalline rods) and the latter from µSR and ac-susceptibility data. Although caution is required because the samples are not high purity single crystals, these results are in clear disagreement with the BCS model.

The behaviour of the dc resistivity in the normal state also appears to be somewhat controversial. It is found that the temperature dependent term is rather different for samples with the same $T_c$ onset value [2,21-26]. Taking into account the phonon-mediated nature of the superconductivity in MgB$_2$, this suggests that some samples appear to contain an additional strong and temperature dependent scattering mechanism.

In the present work, we present the temperature dependence of the microwave surface resistance $R_s$ of separated MgB$_2$ grains. Each grain represents a well connected, although not single crystalline sample. The flux pinning properties and critical current densities of these grains are presented elsewhere [13]. From the value of the surface resistance at $T_c$ we estimate that the dc resistivity of our samples is $\rho_{dc}=25\mu\Omega$cm. This is a consistency check for the microwave measurement and the value is reasonable. We find that $R_s$ is a linear function of temperature below 25K. The absolute value of the surface resistance and the implication for the temperature dependence are discussed below.

## EXPERIMENTAL DETAILS

### Preparation of the sample

Commercially available $MgB_2$ powder was used (Alfa Aesar Co., 98% purity). From this powder three different samples were made. Sample 1 was made of 20 grains each about 200μm diameter, displaced in a commercial optical clean liquid (Opticlean) and placed on the dielectric puck resonator in a symmetric fashion about the central axis. The second and third samples consisted of 50 and 500 grains, respectively. These grains (which were approximately 50μm diameter) were first cleaned ultrasonically in ethanol and fixed to the dielectric puck with wax. In sample number 3, layers of wax were created, with each layer containing some hundred grains. To ensure spatial separation of the grains, the correct position of each grain was checked with an optical microscope.

To eliminate artefacts, it is important to ensure the highest possible purity and smoothness of the surface of each grain and the absence of weak coupling between grains. The grains were examined using a scanning electron microscope before and after cleaning. The ultrasonic cleaning proved to be an effective way of removing submicron size particles from the surface of the grains, as shown in figure 1.

### Experimental setup

Measurements of $R_s(T)$ were made using a $TiO_2$ dielectric puck resonator at 3GHz. The fundamental $TE_{011}$ mode was used for all measurements. The resonator with a tuneable copper cavity was placed on the cold finger of the closed cycle cooler with operation range of temperature 10-300K. Further details are given elsewhere [27].

To obtain the absolute value of the surface resistance, the calculation of the geometric factor was performed using MAFIA (commercially available software)[28].

## RESULTS AND DISCUSSION

Figure 2 shows $R_s(T)$ of the samples before and after cleaning the surface. A quite pronounced difference is observed by cleaning and removing the submicron size particles. The value of the surface resistance drops from $R_s(25K)=1.2m\Omega$ to 0.5mΩ and the temperature dependence becomes less pronounced. The similar behaviour of $R_s(T)$ for sample 2 and 3 confirms the absence of weak links between grains of neighbouring layers in sample 3.

The surface resistance $R_s$ at $T_c$ is 58 mΩ. This can be converted to a dc resistivity $\rho(T_c)$ of 25 μΩcm. This value is consistent with resistivity data obtained on $MgB_2$ polycrystalline pellets [2,21] but is approximately two orders of magnitude higher than $\rho(T_c)$ of $MgB_2$ wire [22,23] and approximately ten times higher than the resistivity of $MgB_2$ epitaxial thin film [26]. Unfortunately we can not draw any conclusion about the exponent $n$ of the temperature dependence of resistivity above the critical temperature $\rho(T) \propto T^n$ because of the low quality factor of the dielectric resonator used and its strong temperature dependence above 100K.

At low temperatures T ≤ 25K $R_s(T)$ shows a linear temperature dependence (open symbols on the insert of Figure 2). Extrapolation of $R_s(T)$ yields a residual surface resistance $R_{res}(3GHz) = 230\mu\Omega$. This value is comparable to $R_{res}(3GHz) =173\mu\Omega$ for BSCCO single crystals [29] (where we have converted the data from 10GHz assuming an $\omega^2$ dependence [30]). The same BSCCO single crystal demonstrated a linear, temperature dependence in $\lambda(T)$ also. A linear temperature dependence was confirmed by similar measurements performed on higher quality BSCCO single crystals with a lower $R_{res}(T)$ and a more gradual slope of $R_s(T)$ [30,31]. The microwave data from the poorer quality BSCCO crystal (re-calculated to 3GHZ) [29] is shown in the insert of Figure 2 (solid symbols). The similarity between BSCCO and $MgB_2$ might suggest that the linear temperature dependence of $R_s$ for $MGB_2$, is intrinsic. But we have to note that the value of the surface resistance at $T_c/2$ of our sample is still quite high. It is approximately two orders of magnitude higher than $R_s$ of Nb measured at $T_c/2$ at 3.7GHz, where $R_s(Nb) = 5\mu\Omega$ [32].

Linear temperature dependence of surface resistance at low temperatures contradicts the activated form of the temperature dependence predicted by the BCS and strong coupling models with s-wave pairing and observed on conventional low-temperature superconductors such as Nb [32] and $Nb_3Sn$ [33].

We must emphasis that these experiments were performed on poly-crystalline samples with high resistivity at $T_c$, lacking smooth surfaces and with some intragranular porosity. Hence, the temperature dependence of the surface resistance could easily be associated with weak link effects within each grain [34]. However, it is also possible to explain $R_s(T) \propto T$ in the framework of two-band superconductivity theory with an s-wave order parameter in each band. The justification for this conjecture is that band structure calculations of $MgB_2$ suggests that there are, at least, two types of bands at the Fermi surface. The first one is a heavy hole band, built up of boron σ orbitals. The second one is a broader band with a smaller effective mass, built up mainly of π boron orbitals. Theoretical arguments claim either the origin of the superconductivity lies in the π band [3], or in the σ band [4-6]. Shulga et al., argued that the temperature dependence of the upper critical field is most accurately reproduced when at least two bands are included at the Fermi level [35]. In the framework of two-band Eliashberg theory it is possible to obtain both linear and quadratic temperature dependences of λ(T)[36] as well as a linear temperature dependence of $R_s(T)$[37] by taking into account scattering on magnetic and nonmagnetic impurities in the band with the smaller gap.

In conclusion, the results of measurements of $MgB_2$ polycrystalline grains are reported. Linear temperature dependence of surface resistance is observed. This behaviour can not be explained by BCS theory. To identify whether the observed temperature dependence reflects intrinsic or extrinsic properties of $MgB_2$ the measurements of the temperature dependence of both components of the surface impedance need to be performed, urgently, on high quality films and single crystals.

**Acknowledgements**

The work is supported by the EPSRC GR/M67445, NPL and the Royal Society.

*Fig.1.* Surface quality before (Fig.1a) and after (Fig.1b) ultrasonic cleaning.

*Fig. 2.* Temperature dependence of the surface resistance of sample 1 (solid squares) and samples 2 (open triangles) and 3 (open symbols). Samples 2 and 3 were ultrasonic cleaned. The inset shows temperature dependence of $MgB_2$ (open squares) and BSCCO (solid circles) [29] recalculated for 3GHz, the line is a guide to the eye.}

+ Visiting from the INFM – Dipartimento di Fisica, Universita'di Salerno, Salerno, I-84081, Italy

1. J. Nagamatsu, N. Nakagawa, T. Muranaka, Y. Zenitani, and J. Akimitsu, Nature **410**, 63 (2001).
2. W.N. Kang, C.U. Jung, Kijoon H.P. Kim, Min-Seok Park, S.Y. Lee, Hyeong-Jin Kim, Eun-Mi Choi, Kyung Hee Kim, Mun-Seog Kim, and Sung-Ik Lee, cond-mat/ 0102313 (2001).

3. J. Kortus, I.I. Mazin, K.D. Belashchenko, V.P. Antropov, and L.L. Boyer, cond-mat/ 0101446 (2001).
4. J.E. Hirsch, cond-mat/0102115 (2001).
5 K.D. Belashchenko, M. van Schilfgaarde, and V.P. Antropov, cond-mat/0102290 (2001).
6. J.M. An and W.E. Pickett, cond-mat/0102391 (2001).
7. H. Kotegawa, K. Ishida, Y. Kitaoka, T. Muranaka, and J. Akimitsu, cond-mat/0102334 (2001).
8. S.L. Bud'ko, G. Lapertot, C. Petrovic, C.E. Cunningham, N. Anderson, and P.C. Canfield, Phys. Rev. Lett. **86**, 1877 (2001).
9. R. Osborn, E.A. Goremychkin, A.I. Kolesnikov and D.G. Hinks, cond-mat/0103064 (2001).
10. B. Lorenz, R.L. Meng and C.W. Chu, cond-mat/0102264 (2001).
11. R.K. Kremer, B.J. Gibson and K. Ahn, cond-mat/0102432 (2001).
12. Ch. Walti, E. Felder, C. Degen, G. Wigger, R. Monnier, B. Delley and H.R. Ott, cond-mat/0102522 (2001).
13. Y. Bugoslavsky, G.K. Perkins, X. Qi, L.F. Cohen and A.D. Caplin, accepted to Nature.
14. G. Rubio-Bollinger, H. Suderow, and S. Vieira, cond-mat/0102242 (2001).
15. G. Karapetrov, M. Iavarone, W.K. Kwok, G.W. Crabtree, and D.G. Hinks, cond-mat/ 0102312 (2001).
16. A. Sharoni, I. Felner, and O. Millo, cond-mat/0102325 (2001).
17. H. Schmidt, J.F. Zasadzinski, K.E. Gray, and D.G. Hinks, cond-mat/0102389 (2001).
18. E. Bascones and F. Guinea, cond-mat/0103190.
19. S.L. Li, H.H. Wen, Z.W. Zhao, Y.M. Ni, Z.A. Ren, G.C. Che, H.P. Yang, Z.Y. Liu and Z.X. Zhao, cond-mat/0103032 (2001).
20. C. Panagopoulus, B.D. Rainford, T. Xiang, C.A. Scott, M. Kambara and I.H. Inoue, cond-mat/0103060 (2001).
21. C.U. Jung, Min-Seok Park, W.N. Kang, Mun-Seog Kim, S.Y. Lee and Sung-Ik Lee, cond-mat/0102215 (2001).
22. P.C. Canfield, D.K. Finnemore, S.L. Bud'ko, J.E. Ostenson, G. Lapertot, C.E. Cunningham and C. Petrovich, cond-mat/0102289 (2001).
23. S.L. Bud'ko, C. Petrovich, G. Laperot, C.E. Cunningham and P.C. Canfield, cond-mat/0102413 (2001).
24. B. Gorshunov, C.A. Kuntscher, P. Haas, M. Dressel, F.P. Mena, A.B. Kuz'menko, D. van der Marel, T. Muranaka, J. Akimitsu, cond-mat/0103164 (2001).
25. Jai Seok Ahn and Eun Jip Choi, cond-mat/0103169 (2001).
26. W.N. Kang, Hyeong-Jin Kim, Eun-Mi Choi, Chang Uk Jung and Sung-Ik Lee, cond-mat/0103179 (2001).
27. N. McN Alford et al., J. of Supercond. **10**, 467 (1997)
28. L. Hao, J. Gallop, A. Purnell, L. Cohen and S. Thiess, J. Supercond (2000).
29. T. Jacobs, S. Sridhar, Q. Li, G.D. Gu and N. Koshizuka, Phys. Rev. Lett. **75**, 4516 (1995).
30. S.-F. Lee, D.C. Morgan, R.J. Ormeno, D.M. Broun, R.A. Doyle, J.R. Waldram, and K. Kadowaki, Phys. Rev. Lett. **77**, 735 (1996).
31. D.V. Shovkun, M.R. Trunin, A.A. Zhukov, Yu.A. Nefyodov, N. Bontemps, H. Enríquez, A. Buzdin, M. Daumens, and T. Tamegai, JETP Letters **71**, 92 (2000).
32. J. Halbritter, Z.Phys. **266**, 209 (1974).
33. H. Piel, Nuclear Methods and Physics Research **A287**, 294 (1990).
34. T. Jacobs, K. Numssen, R. Schwab, R. Heidinger, J. Halbritter, IEEE Trans. Of Appl. Supercond. **7**, 1917 (1997).
35. S.V. Shulga, S.-L. Drechsler, H. Eschrig, H. Rosner and W. Pickett, cond-mat/0103154, (2001).
36. A.A. Golubov, M.R. Trunin, A.A. Zhukov, O.V. Dolgov, and S.V. Shulga, Pis'ma v Zh. Exp. Teor. Fiz. **62**, 477-482 (1995) [JETP Lett. **62**, 42-48 (1995)].
37. A.A. Golubov, M.R. Trunin, A.A. Zhukov, O.V. Dolgov, and S.V. Shulga, J.Phys. I France **6**, 2275-2290 (1996).

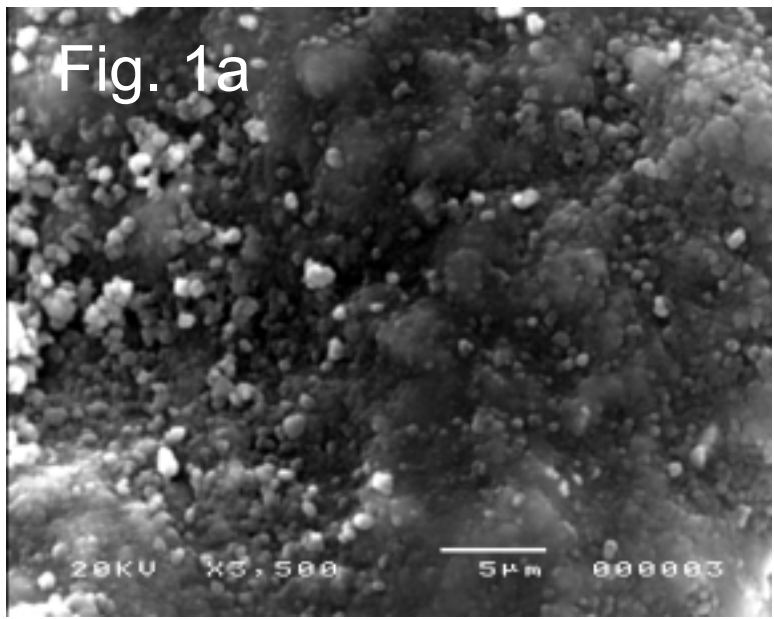

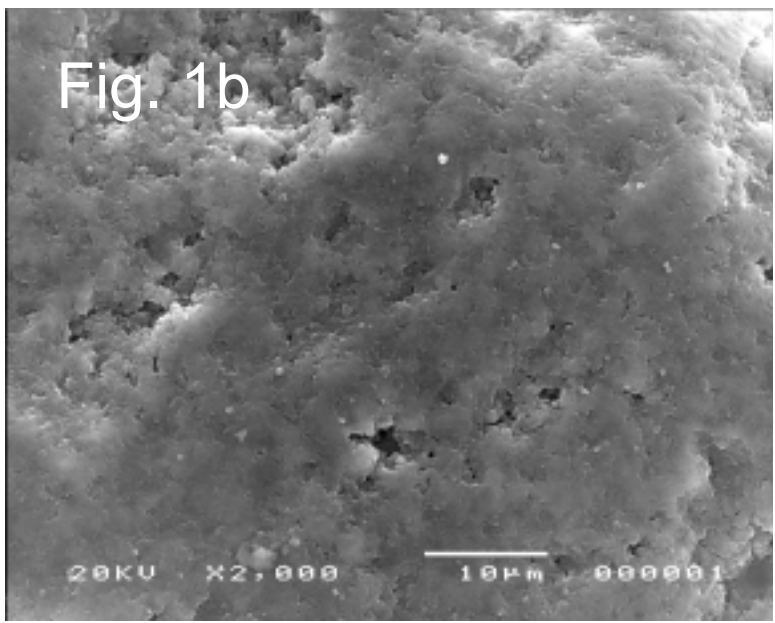

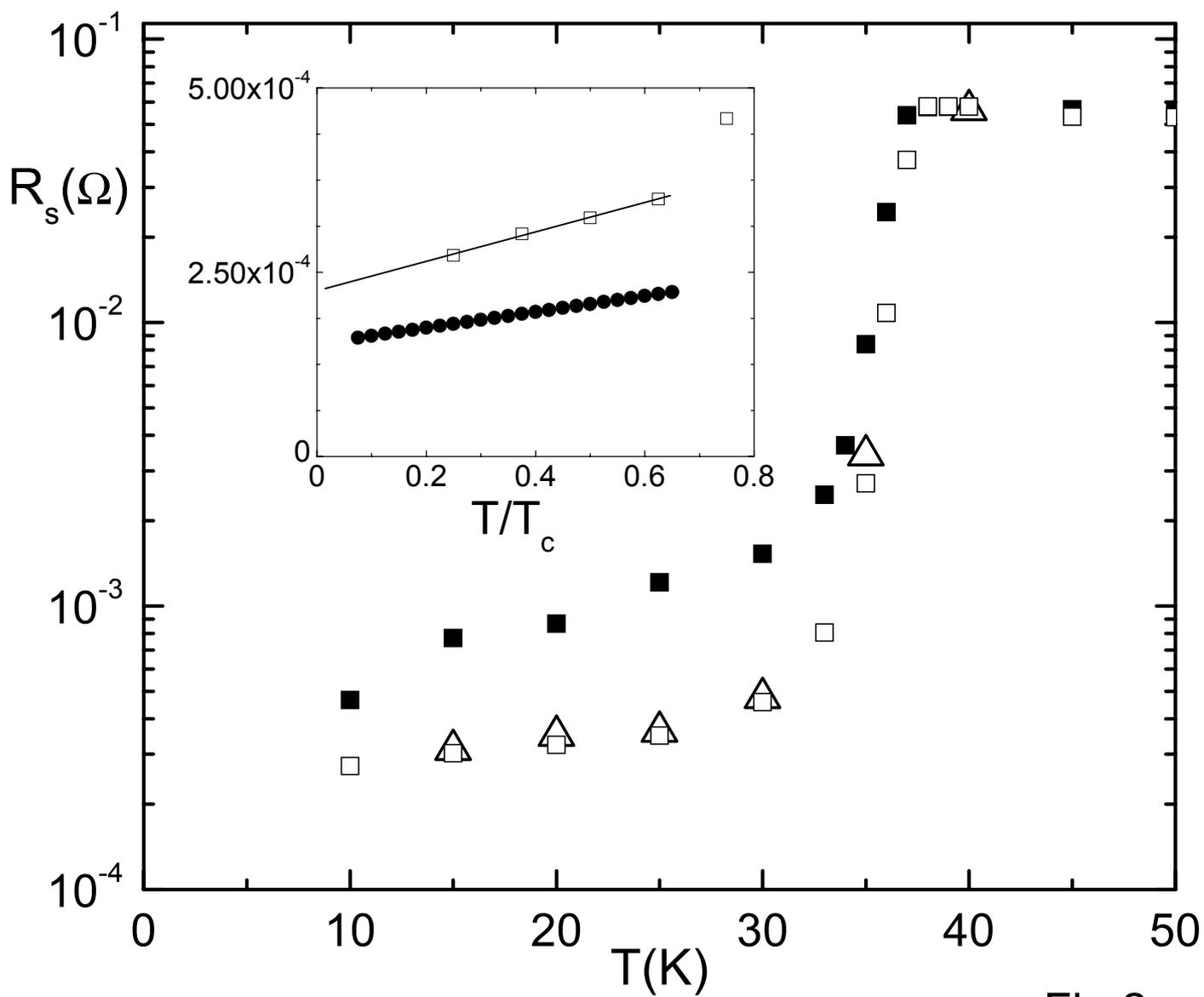

Fig 2.